\documentclass[12pt]{article}
\usepackage[dvips]{graphicx}
\usepackage{epsfig,amsmath}
\begin{document}
\thispagestyle{empty}
\begin{center}

{\bf THE RELEVANCE OF POSITIVITY IN SPIN PHYSICS \footnote{Presented by J. Soffer at CIPANP 2009, San Diego California, USA, May 26-31, 2009}}\\

\vskip1.4cm
{\bf Jacques Soffer}
\vskip 0.3cm
Physics Department, Temple University, Philadelphia, PA 19122-6082, USA\\
\vskip 0.5cm
{\bf Xavier Artru}
\vskip 0.3cm
Universit\'e de Lyon, IPNL and CNRS/IN2P3, 69622 Villeurbanne, France
\vskip 0.5cm
{\bf Mokhtar Elchikh}
\vskip 0.3cm
Universit\'e des Sciences et de Technologie d'Oran, El Menauoer, Oran, Algeria
\vskip 0.5cm
{\bf Jean-Marc Richard}
\vskip 0.3cm
LPSC, Universit\'e  Joseph Fourier, CNRS/IN2P3, INPG, Grenoble, France
\vskip 0.5cm
{\bf Oleg Teryaev}
\vskip 0.3cm
Bogoliubov Laboratory of Theoretical Physics, JINR, 141980 Dubna, Moscou region, Russia
\vskip 1.5cm
{\bf Abstract}\end{center}
Positivity reduces substantially the allowed domain for spin observables. We briefly
recall some methods used to determine these domains and
give some typical examples for exclusive and inclusive spin-dependent reactions.

\vskip 0.5cm

\noindent {\it Key words}: constraints, positivity domains, spin observables\\
\noindent PACS numbers: 13.75.-n, 13.85.-t, 13.88.+e

\vskip 0.5cm

\section{INTRODUCTION}
Spin observables for any particle reaction, contain some unique information which allow a deeper understanding of the
nature of the underlying dynamics and this is very usefull to check the validity of theoretical assumptions. We 
emphasize the relevance of positivity in spin physics, which puts non-trivial model independent constraints on spin
observables. If one, two or several observables are measured, the constraints can help to decide which new observable will provide
the best improvement of knowledge. Different methods can be used to establish these constraints and they have been presented together with
many interesting cases in a recent review article \cite{aerst}.\\
If X, Y, Z are spin observables with the standard normalization $-1\leq X\leq +1$, the
domain for the pair of observables (X,Y) is often {\it smaller} than the square $[-1,+1]^2$ and
the domain for the triple of observables (X,Y,Z) is often {\it smaller} than the cube $[-1,+1]^3$. Explicit 
inequalities are obtained, relating two or three spin observables, for instance
$X^2+Y^2 \leq 1$, corresponding to a disk, or $X^2+Y^2+Z^2 \leq 1$, corresponding to a sphere, but one is also led to triangles, tetrahedrons, etc...(see below).
\section{EXCLUSIVE REACTIONS}
{\it i) - $\pi N \to \pi N$ scattering}\\
A scattering process involving spinning particles and described by $n$ {\it complex} amplitudes
is fully determined by $(2n-1)$ {\it real} functions, up to an over-all phase. Since
there are $n^2$ possible measurements for this reaction, we must have $(n-1)^2$ independent
quadratic relations between the $n^2$ observables.
First consider the
simplest case of an exclusive two-body reaction with a spin 0 and a spin 1/2 particle, namely
$0 + 1/2 \to 0 + 1/2$. As an example $\pi N \to \pi N$ is described in terms of two amplitudes, the
non-flip $f_+$ and the flip $f_-$, so we have four observables, the
cross section $d\sigma/dt$, the polarization $P_n$ and two rotation parameters $R$ and $A$. There is
one well known {\it quadratic} relation, that is $P_n^2 + A^2 + R^2 = 1$.
If all three are measured they should lie on the surface of the unit sphere, which is a consistency check of the data.\\
Concerning the crossed reaction $\bar pp \to \pi\pi$, large spin effects have been seen at LEAR. One has an analogous identity as the one above for $\pi N$, namely $A_n^2+A^2_{mm}+A^2_{ml}=1$. Hence $|A_n|=1$, implies $A_{mm}=A_{ml}=0$. For $\bar pp \to \bar KK$ similar results were obtained \cite{aerst}.

{\it ii) - $ Spin 1/2 + Spin 1/2 \to Spin 1/2 + Spin 1/2$}\\
For the reaction $1/2 + 1/2 \to 1/2 + 1/2$, for example $pp$ elastic scattering, we have five amplitudes, therefore
twenty-five observables and sixteen quadratic relations between them. For the derivation
one considers a 5x5 matrix of the observables, which is positive Hermitian
and the final results can be found in Ref.~\cite{bs}. These relations are very useful to check
the data and when one observable is not measured, it is set to zero and the equality becomes an
inequality. In particular one finds the following very simple condition $A^2_{LL} + D^2_{NN} \leq 1$,
between the two-spin correlation parameters $A_{LL}$ and $D_{NN}$. 
This condition turns out to be very usefull because if 
one of the parameters is close to one, the other one is bounded to be near 
zero. In the reaction $\bar p p \to \bar \Lambda \Lambda$, one
has found in a certain kinematic region $A_{LL}=-1$, so before making any measurement 
one can conclude that $D_{NN} \sim 0$, which is also a no-go theorem for some theoretical
considerations \cite{jmr}.\\
An empirical method consists of generating randomly six complex amplitudes, for $\bar pp \to \bar \Lambda \Lambda$, to compute the spin
observables and look at the constraints (see Fig.~1). The empirical method was extended to the case of triple of spin observables.
Several remarkable shapes were discovered for the allowed domain:
sphere, cylinder, cone, pyramid, tetrahedron, octahedron, intersection of two cylinders, 
intersection of three orthogonal cylinders, coffee filter, etc...\cite{aerst}.

{\it iii) - Pseudoscalar meson photoproduction}\\
Consider the reactions $\gamma N \to K Y$, with $Y=\Lambda, \Sigma$, where the incoming
photon beam is polarized, the nucleon target is polarized and the polarization of the 
outgoing $\Lambda, \Sigma$ is measured. Need to determine four complex amplitudes, {\it i.e.} seven real numbers. One
can perform sixteen different experiments: the unpolarized cross section, three single spin asymmetries, the linearly
polarized photon asymmetry $\Sigma^{\gamma}$, the polarized target asymmetry $A_N$ and the recoil baryon
polarization $P_Y$. We have the
linear relations $|A_N \pm P_Y|\leq 1 \pm \Sigma^{\gamma}$, which give the tetrahedron shown in Fig. 2. The 
allowed volume is 1/3 of the entire cube $[-1,+1]^3$.
Finally, there are four double correlations between the target
and the recoil baryon spins in the scattering plane with unpolarized photons, four
double spin correlations with linearly polarized photons and four double spin correlations
with circularly polarized photons.
Clearly, these sixteen measurements must be constrained by nine quadratic relations, for example, $P_Y$ and the double 
correlation parameters between the circularly polarized photon and the recoil
baryon spin along the directions $\hat x$ and $\hat z$ in the scattering plane, $C^Y_x$ and $C^Y_z$, i.e.
$(P_Y)^2 + (C^Y_x)^2 + (C^Y_z)^2 \leq 1$.  Following the analysis of the CLAS data at Jefferson Lab., it is almost
saturated and this implies the saturation of other quadratic constraints with three or more observables \cite{aerst}.

\section{INCLUSIVE REACTIONS}
{\it i) - Asymmetries in DIS}\\
A bound on the DIS transverse asymmetry $A_2$ has been established
long time ago and reads $|A_2| \leq \sqrt{R}$
where $R$ is the standard ratio $\sigma_L/\sigma_T$.
There is an improved version of this positivity constraint, namely,
$|A_2| \leq \sqrt{R(1+A_1)/2}$, which is very relevant when the DIS longitudinal asymmetry $A_1$ is negative. This
is the case for $A_1^n$, with a neutron target in a certain kinematic range.\\

{\it ii) - Spin-transfer observables}\\
Consider a parity conserving inclusive reaction of the type, $a(spin 1/2) + b(unpol.) \to c(spin 1/2) + X$.
One can define {\it eight} observables, which must satisfy \footnote{ {\it NOTE}: The eight transverse momentum dependent quark distributions obey the same constraints since they are related to $nucleon(p,S) \to quark(k,S^{'}) + X$.}
\begin{equation}
(1 \pm D_{NN})^2  \geq (P_{cN} \pm A_{aN})^2 + (D_{LL} \pm D_{SS})^2 + (D_{LS} \mp D_{SL})^2
\end{equation}
If we concentrate for the moment on the
case where the particle spins are {\it normal} to the scattering, 
for example for $p^{\uparrow} p \to \Lambda^{\uparrow} X$, one has $1 \pm D_{NN}  \geq  | P_{\Lambda} \pm A_N |$. The
corresponding allowed domains are displayed on Fig.~3.\\

{\it iii) - Quark Transversity Distribution $\delta q(x,Q^2)$}\\
In addition to $q(x,Q^2)$ and $\Delta q(x,Q^2)$, there is a new distribution function $\delta q(x,Q^2)$, chiral odd, leading twist,
which decouples from DIS and has been indirectly extracted recently for the first time.
It must satisfy the following positivity bound \cite{js}, $q(x,Q^2) + \Delta q(x,Q^2) \geq 2|\delta q(x,Q^2)|$, which 
survives up to NLO corrections. The corresponding allowed triangle occured also for $\bar pp\to\bar \Lambda\Lambda$, as seen above.\\ 
\newpage
{\bf Acknowledgements} J.S. is very grateful to the organizers of the tenth Conference
on the Intersections of Particle and Nuclear Physics, CIPANP 2009, and in particular to Marvin
Marshak, for giving him the opportunity to present this work and for some financial
support.

\begin{figure}[p]
\hspace*{-2cm}
  \begin{minipage}{5.0cm}
  \epsfig{figure=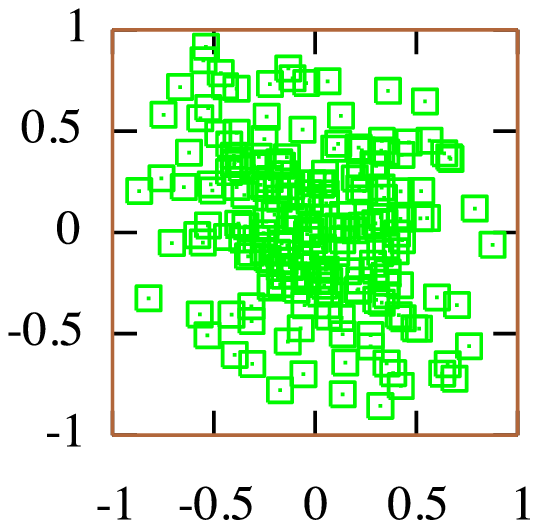,height=3.5cm}
  \end{minipage}
  \begin{minipage}{5.0cm}
  \epsfig{figure=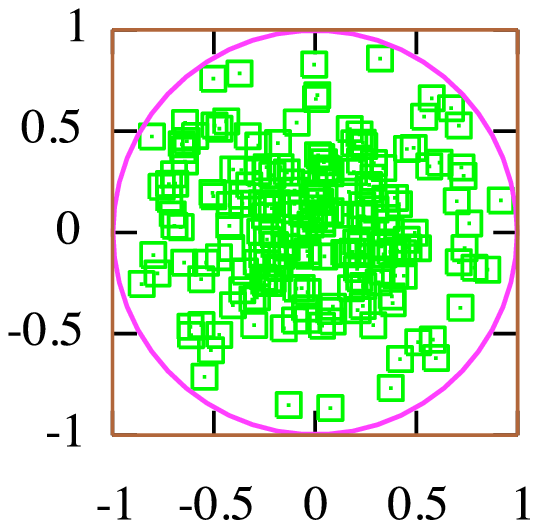,height=3.5cm}
  \end{minipage}
    \begin{minipage}{5.0cm}
  \epsfig{figure=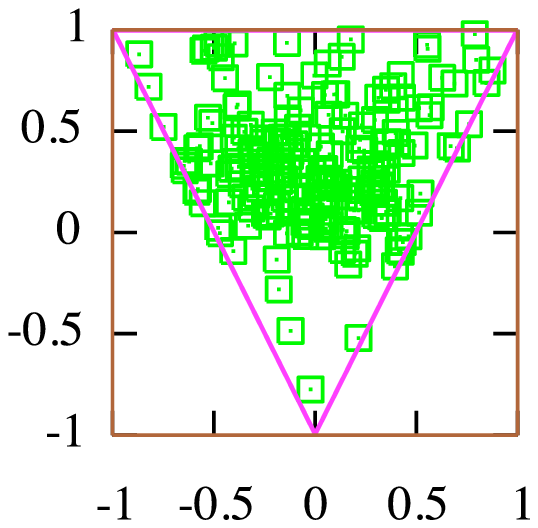,height=3.5cm}
  \end{minipage}\\
\caption{
Random simulation for $\bar pp \to \bar \Lambda \Lambda$ of $P_n$ versus $A_n$ (square), $A_n$ versus $D_{mm}$ (disk) and $P_n$ versus $C_{nn}$ (triangle) .}
\label{fi:fig1}
\vspace*{-2.5ex}
\end{figure}

\begin{figure}[p]
 \epsfig{figure=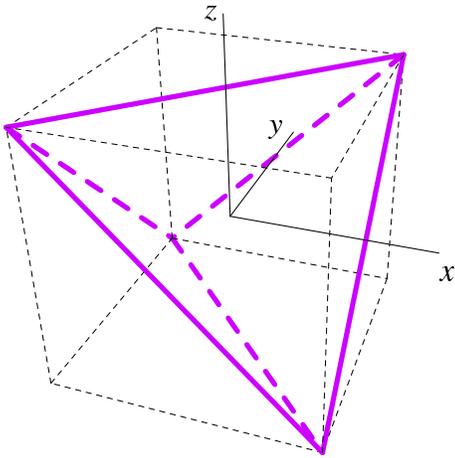,width=6.0cm}
\caption{
Tetrahedron domain limited by inequalities $|A_N \pm P_Y|\leq 1 \pm \Sigma^{\gamma}$, for the photoproduction observables $x=A_N$, $y=P_Y$ and $z=\Sigma^{\gamma}$.}
\label{fi:fig2}
\end{figure}
\newpage
\begin{figure}[p]
\begin{center}
 \epsfig{figure=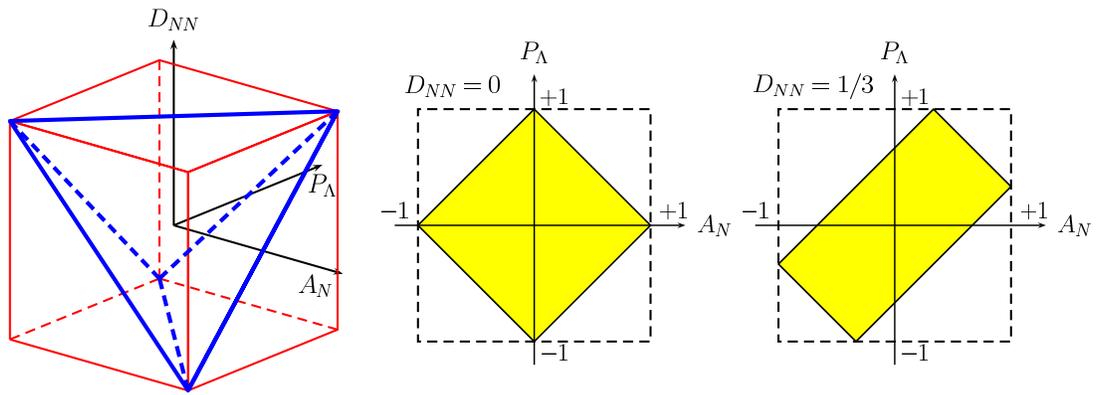,angle=-90,width=12.0cm}
 \end{center}
 \vspace*{2.5cm}
\caption{The allowed domain corresponding to the constraints Eq.~(1) (left). The slice of the full domain
for $D_{NN}=0$ (middle) and for $D_{NN}=1/3$ (right).}
\label{fi:fig3}
\end{figure}

\end{document}